\documentstyle[12pt]{article}

\def\op#1{\mathop{\fam0 #1}\limits}

\newcommand{\Id}{{\rm Id\,}}

\newcommand{\dif}{{\rm Diff\,}}
\newcommand{\lng}{\langle}
\newcommand{\rng}{\rangle}

\newcommand{\beq}{\begin{equation}}
\newcommand{\eeq}{\end{equation}}
\newcommand{\ben}{\begin{eqnarray}}
\newcommand{\een}{\end{eqnarray}}
\newcommand{\be}{\begin{eqnarray*}}
\newcommand{\ee}{\end{eqnarray*}}
\newcommand{\bea}{\begin{eqalph}}
\newcommand{\eea}{\end{eqalph}}

\newcommand{\cH}{{\cal H}}

\newcommand{\cS}{{\cal S}}

\newcommand{\bR}{{\bf R}}

\newcommand{\al}{\alpha}

\newcommand{\la}{\lambda}

\newcommand{\f}{\phi}

\newcommand{\m}{\mu}

\newcommand{\si}{\sigma}
\newcommand{\Si}{\Sigma}

\newcommand{\wt}{\widetilde}
\newcommand{\wh}{\widehat}

\newcommand{\dr}{\partial}

\newcommand{\ot}{\otimes}

\let\ssection=\section
\renewcommand{\section}{\setcounter{equation}{0}\ssection}

\newcounter{eqalph}
\newcounter{equationa}
\newcounter{remark}
\newcounter{example}
\newcounter{theorem}
\newcounter{proposition}
\newcounter{lemma}
\newcounter{corollary}
\newcounter{definition}
\newenvironment{eqalph}{\stepcounter{equation}
\setcounter{equationa}{\value{equation}}
\setcounter{equation}{0}

\begin{eqnarray}}{\end{eqnarray}\setcounter{equation}{\value{equationa}}}

\def\theremark{\arabic{remark}}
\def\therexample{\arabic{remark}}

\def\thedefinition{\arabic{definition}}

\newenvironment{rem}{\refstepcounter{remark}\medskip\noindent{\it Remark
\theremark:}}{\medskip}

\newenvironment{prop}{\refstepcounter{definition} \medskip
\noindent{\it Proposition \thedefinition:}}{\medskip}

\newenvironment{defi}{\refstepcounter{definition} \medskip
\noindent{\it Definition \thedefinition:}}{\medskip}

\begin{document}
\hbox{}

{\parindent=0pt

{\large\bf Quantum mechanics with time-dependent parameters}
\bigskip

{\sc G. Sardanashvily}\footnote{Electronic mail:
sard@grav.phys.msu.su; sard@campus.unicam.it}

{\sl Department of Theoretical Physics, 
Moscow State University, 117234 Moscow, Russia}
\bigskip\bigskip

{\small

{\bf Abstract}. Composite bundles $Q\to\Si\to \bR$, where $\Si\to\bR$ is
the parameter bundle, provide the adequate mathematical description of
classical mechanics with time-dependent parameters. We show that the Berry's
phase phenomenon is described in terms of connections on composite Hilbert
space bundles. }
}
\bigskip\bigskip

\noindent
{\bf I.}
\bigskip

Smooth fiber bundles $Q\to\bR$ over a time axis $\bR$ provide the adequate
formulation of classical time-dependent mechanics treated as a
particular field theory \cite{sard98,book98}. Let us consider a 
mechanical system 
depending on time-dependent parameters. These parameters can be seen as
sections of some smooth fiber bundle $\Si\to\bR$. Then the
configuration space of a mechanical system with time-dependent
parameters can be seen as the composite fiber bundle
\beq
Q\to\Si\to\bR. \label{z500'}
\eeq
In classical mechanics $Q\to\Si$ is a smooth finite-dimensional fiber bundle.
In quantum mechanics $Q\to\Si$ is a $C^*$-algebra fiber bundle or a Hilbert
space fiber bundle \cite{book00}.

The following two facts make the composite fiber bundle (\ref{z500'}) useful
for our purpose.

(i) Given a section $h$ of a parameter bundle $\Si\to\bR$, the
pull-back bundle $h^*Q$ over $\bR$ describes a mechanical system under 
the fixed parameter functions $h(t)$.

(ii) Given a connection $A_\Si$ on the fiber bundle $Q\to\Si$, the
pull-back connection $h^*A_\Si$ on the pull-back bundle $h^*Q\to\bR$
depends in a certain way on the parameter functions $h(t)$, and
characterizes the dynamics of a mechanical system with time-dependent
parameters.

This work is devoted to quantum mechanics with classical
parameters where connections on composite Hilbert
space bundles play the role of Berry connections.
\bigskip

\noindent
{\bf II.}
\bigskip

Recall that by a smooth composite bundle is meant the composition of
fiber bundles
\beq
Y\to \Si\to X, \label{1.34}
\eeq
where $\pi_{Y\Si}: Y\to\Si$ and $\pi_{\Si X}: \Si\to X$
are smooth fiber bundles \cite{book00,book}. It is
provided with an atlas of fibered coordinates $(x^\la,\si^m,y^i)$, where
$(x^\m,\si^m)$ are fibered coordinates on the fiber bundle $\Si\to X$ and the
transition functions $\si^m\to\si'^m(x^\la,\si^k)$ are independent of the
fiber coordinates $y^i$.

\begin{prop} \label{1}
Given a composite fiber bundle (\ref{1.34}), let $h$ be a global section
of the fiber bundle $\Si\to X$. Then the restriction
\beq
Y_h=h^*Y \label{S10}
\eeq
of the fiber bundle $Y\to\Si$ to $h(X)\subset \Si$ is a subbundle 
$i_h: Y_h\hookrightarrow Y$
of the fiber bundle $Y\to X$. 
\end{prop}

Let us consider a connection
\beq
A_\Si=dx^\la\ot (\dr_\la + A_\la^i\dr_i) +d\si^m\ot (\dr_m + A_m^i\dr_i) 
: Y\to J^1_\Si Y
\label{b1.113}
\eeq
on the fiber bundle $Y\to \Si$. Given a section
$h$ the fiber bundle $\Si\to X$, the
connection $A_\Si$ (\ref{b1.113}) induces the pull-back
connection
\beq
A_h=i_h^*A_\Si=dx^\la\ot[\dr_\la+((A^i_m\circ h)\dr_\la h^m
+(A\circ h)^i_\la)\dr_i] \label{mos83}
\eeq
on the pull-back bundle $Y_h$ (\ref{S10}). 

Note that, in quantum theory,
one follows the notion of a connection phrased in algebraic terms as a
connection on modules in comparison with the pure geometric one in
classical theory. Here, we restrict our consideration to
connecions on modules over the ring $C^\infty(X)$ of smooth real
functions on a manifold $X$ \cite{book00,kosz60}.

\begin{defi} \label{mos086} 
A connection on a
$C^\infty(X)$-module $\cS$ assigns to each vector field $\tau$ on a
manifold $X$ an $\cS$-valued first order differential operator
$\nabla_\tau\in \dif_1(\cS,\cS)$ on $\cS$ which 
obeys the Leibniz rule
\beq
\nabla_\tau (fs)= (\tau\rfloor df)s+ f\nabla_\tau s,\quad f\in
C^\infty(X), \quad s\in\cS. \label{mos087}
\eeq
\end{defi}

If $\cS$ is a module of global sections of a smooth 
vector bundle $Y\to X$ over a manifold $X$, Definition \ref{mos086} is
equivalent to the familiar geometric definition of a connection on
$Y\to X$.

\bigskip 

\noindent 
{\bf III.}
\bigskip

Let us consider a quantum mechanical systems depending on a finite
number of real 
classical parameters given by sections of a smooth parameter bundle $\Si\to
\bR$. For the sake of simplicity, we fix a trivialization
$\Si=\bR\times Z$, coordinated by $(t,\si^m)$. Although it may happen that the
parameter bundle $\Si\to \bR$ has no preferable trivialization, e.g., if
one of parameters is a velocity of a reference frame.

Recall that, in the framework of  algebraic quantum theory, a quantum
system is characterized by a
$C^*$-algebra
$A$ and a positive (hence, continuous) form $\f$ on $A$ which defines the
 representation $\pi_\f$ of
$A$ in a Hilbert space $E_\f$ with a cyclic vector $\xi_\f$ such that 
\be
\f(a)=\lng \pi_\f(a)\xi_\f|\xi_\f\rng, \qquad \forall a\in A.
\ee
One says that
$\f(a)$ is a mean value of the operator $a$ in the state $\xi_\f$. 

It should be emphasized that, in quantum mechanics, 
a time also plays the role of a classical parameter. Indeed, 
all relations between
operators in quantum mechanics are simultaneous, while a computation of a mean
value of an operator in a quantum state  does not imply an integration over a
time. It follows that, at each moment, we have a quantum system, but these
quantum systems are different at different instants. 
Though they may be isomorphic to each other.
This characteristic is extended to other
classical parameters. Namely, we assign a
$C^*$-algebra
$A_\si$ to each point
$\si\in \Si$ of the parameter bundle $\Si$, and treat
$A_\si$ as a quantum system under fixed values $(t,\si^m)$ of the parameters. 

\begin{rem} \label{+329} 
Let us emphasize that one should distinguish classical parameters from
coordinates which a wave function can depend on. 
Let $\{A_q\}$ be a set of $C^*$-algebras parameterized by points of a
 locally compact topological space $Q$. Let all $C^*$-algebras
$A_q$ are isomorphic to each other and to some $C^*$-algebra $A$. We
consider a locally trivial topological fiber bundle $P\to Q$ whose typical
fiber is the $C^*$-algebra $A$, i.e., transition functions of this fiber
bundle provide automorphisms of $A$. The set $P(Q)$ of continuous sections of
this fiber bundle is a *-algebra with respect to fiberwise
operations. Let us consider a
subalgebra $A(Q)\subset P(Q)$ which consists of sections $\al$ of $P\to Q$ 
such that the real function
$||\al(q)||$ vanishes at infinity of $Q$. For $\al\in A(Q)$, put
\be
||\al||=\op{\rm sup}_{q\in Q}||\al(q)||<\infty.
\ee
With this norm, $A(Q)$ is a $C^*$-algebra \cite{dixm}.
One can consider a quantum system characterized by this  $C^*$-algebra.
In this case, elements of the set $Q$ are not classical parameters as
follows. Given an element $q\in Q$, the assignment 
\beq
A(Q)\ni \al\mapsto \al(q)\in A \label{+325}
\eeq
is a $C^*$-algebra epimorphism. Let $\pi$ be a representation of $A$. 
Then the assignment (\ref{+325}) yields a representation $\rho(\pi,q)$ of the
$C^*$-algebra
$A(Q)$. If $\pi$ is an irreducible representation of the $C^*$-algebra $A$, 
then $\rho(\pi,q)$ is an irreducible representation of $A(Q)$. Moreover,
the irreducible representations $\rho(\pi,q)$ and $\rho(\pi,q')$
of $A(Q)$ are not equivalent \cite{dixm}. Therefore there is one-to-one
correspondence (but not a 
homeomorphism) between the spectrum $\wh{A(Q)}$ of the $C^*$-algebra
$A(Q)$ and the product $Q\times\wh A$ of $Q$ and the spectrum $\wh A$
of the $C^*$-algebra $A$. It follows that one can find
representations of the $C^*$-algebra $A(Q)$ among direct integrals of
representations of $A$ with respect to some measure on $Q$. Let $\m$ be a
positive measure of total mass 1 on the locally compact space $Q$ ,
and let $\f$ be a positive form on $A$. Then the function $q\mapsto
\f(\al(q))$, $\forall \al\in A(Q)$, is a $\m$-measurable, while the
integral
\be
\f(\al)=\int \f(\al(q))\m(q) 
\ee
provides a positive form on the $C^*$-algebra $A(Q)$. Roughly speaking, a
computation of a mean value of an operator $\al\in A(Q)$ implies an
integration with respect to some measure on $Q$ in general. This is not the
case of quantum systems depending on classical parameters $q\in Q$.
\end{rem} 

We simplify our consideration in order to single out the manifested
Berry's phase phenomenon. 
Let us assume that all algebras
$C^*$-algebras $A_\si$, $\si\in\Si$, are isomorphic to the von Neumann
algebra $B(E)$ of bounded operators 
in some Hilbert space
$E$, and consider a
locally trivial Hilbert space bundle $\Pi\to \Si$ with the typical fiber
$E$ and smooth transition functions \cite{muj}.
Smooth sections of  $\Pi\to \Si$ constitute a module
$\Pi(\Si)$ over the ring $C^\infty(\Si)$ of real functions on $\Si$. 
In accordance with Definition \ref{mos086}, a connection $\wt\nabla$  on
$\Pi(\Si)$ assigns to each vector field $\tau$ on $\Si$ a first order
differential operator 
\beq
\wt \nabla_\tau\in
\dif_1(\Pi(\Si),\Pi(\Si))
\label{+320}
\eeq
which obeys the Leibniz rule
\be
\wt\nabla_\tau (fs)= (\tau\rfloor df)s+ f\wt\nabla_\tau s, \qquad
s\in
\Pi(\Si),
\qquad f\in C^\infty(\Si).
\ee
Let $\tau$ be a vector field on $\Si$ such that $dt\rfloor\tau=1$.
Given a trivialization chart of the Hilbert space bundle $\Pi\to \Si$, the
operator
$\wt\nabla_\tau$ (\ref{+320}) reads
\beq
\wt\nabla_\tau(s) =(\dr_t - i\cH(t,\si^i)) s  +\tau^m(\dr_m
- i\wh A_m(t,\si^i))s,
\label{+321} 
\eeq
where $\cH(t,\si^i)$, $\wh A_m(t,\si^i)$ for each $\si\in\Si$ are bounded
self-adjoint operators in the Hilbert space $E$. 

Let us consider the composite fiber bundle $\Pi\to \Si\to\bR$.
Similarly to the case of smooth composite fiber bundles (see Proposition
\ref{1}), every section $h(t)$ of the parameter bundle $\Si\to\bR$ defines
the subbundle $\Pi_h=h^*\Pi\to\bR$ of the composite fiber bundle $\Pi\to
\bR$ whose typical fiber is the Hilbert space $E$. Accordingly, the connection
$\wt\nabla$ 
(\ref{+321}) on the $C^\infty(\Si)$-module $\Pi(\Si)$ defines the pull-back
connection
\beq
\nabla_h(\psi) = [\dr_t - i(\wh A_m(t,h^i(t))\dr_t h^m+\cH(t,h^i(t))]\psi
\label{+322} 
\eeq  
on the $C^\infty(\bR)$-module $\Pi_h(\bR)$ of sections $\psi$ of the
Hilbert space bundle $\Pi_h\to\bR$.

As in the case of smooth fiber bundles, we say that a section $\psi$ of the
fiber bundle 
$\Pi_h\to\bR$ is an integral section of the connection (\ref{+322}) if 
\beq
\nabla_h(\psi) =[\dr_t -i(\wh A_m(t,h^i(t)) \dr_t
h^m+\cH(t,h^i(t))]\psi=0.
\label{+323}
\eeq
One can think of the equation (\ref{+323}) as being the Shr\"odinger equation
for a quantum system depending on the parameter function $h(t)$. Its
solutions take the form
\beq
G_t=T\exp\left[ i\op\int_0^t (\wh A_m\dr_{t'}h^m + \cH)dt'\right],
\label{+356}
\eeq
where $G_t$ is the time-ordered exponent. 
The term $i\wh A_m(t,h^i(t)) \dr_t
h^m$ in the Shr\"odinger equation (\ref{+323}) is responsible for the 
Berry's phase phenomenon, while $\cH$ is treated as an ordinary Hamiltonian of
a quantum system.  

To show the 
 Berry's phase phenomenon clearly, 
we simplify again the system under consideration. Given a
trivialization of the fiber bundle $\Pi\to\bR$ and the above mentioned
trivialization 
$\Si=\bR\times Z$ of the parameter bundle $\Si$, let us suppose that the
components $\wh A_m$ of the connection $\wt\nabla$ (\ref{+321}) are independent
of
$t$ and that the operators $\cH(\si)$ commute with the operators $\wh
A_m(\si)$ at all points of the curve $h(t)\subset \Si$.  Then the operator
$G_t$ (\ref{+356}) takes the form 
\beq
G_t=T\exp\left[ i\op\int_{h([0,t])} \wh A_m(\si^i) d\si^m \right]\cdot 
T\exp\left[ i\op\int_0^t\cH(t')dt'\right]. \label{+356'}
\eeq
One can think of the first factor in the right-hand side of the expression
(\ref{+356'}) as being the operator of a parallel transport along the curve
$h([0,t])\subset Z$ with respect to the pull-back connection 
\beq
\nabla= i^*\wt\nabla = \dr_m - i\wh A_m(t,\si^i) \label{+357}
\eeq
 on
the fiber bundle $\Pi\to Z$, defined by the imbedding 
\be
i: Z\hookrightarrow
\{0\}\times Z\subset \Si.
\ee
Note that, since operators $\wh A_m$ are independent of time, one can
utilize any 
imbedding of $Z$ to $\{t\}\times Z$. 

Moreover, the connection $\nabla$
(\ref{+357}), called the Berry connection, can
be seen as a connection on some principal fiber bundle
$P\to Z$ for the group $U(E)$ of unitary operators in the Hilbert space $E$.
Let the curve
$h([0,t])$ be closed,  while the holonomy group of the connection
$\nabla$ at the point $h(t)=h(0)$ is not trivial. Then the unitary operator 
\beq
T\exp\left[ i\op\int_{h([0,t])} \wh A_m(\si^i) d\si^m \right] \label{+359}
\eeq 
is not the identity. For example, if 
\beq
i\wh A_m(\si^i)=iA_m(\si^i)\Id_E \label{+360}
\eeq
is a
$U(1)$-principal connection on $Z$, then the operator (\ref{+359}) is 
the well-known Berry phase factor
 \be
\exp\left[ i\op\int_{h([0,t])} A_m(\si^i) d\si^m \right]. 
\ee
If (\ref{+360}) is a curvature-free connection,  Berry's phase 
is exactly the Aharonov--Bohm effect  on the
parameter space $Z$. 

The following variant of the Berry's phase phenomenon leads us to a principal
bundle for familiar finite-dimensional Lie groups. Let $E$ be a separable
Hilbert space which is the Hilbert sum of $n$-dimensional eigenspaces of the
Hamiltonian $\cH(\si)$, i.e.,
\be
E= \op\bigoplus_{k=1}^\infty E_k, \qquad E_k=P_k(E),
\ee
where $P_k$ are the projection operators, i.e.,
\be
H(\si)\circ P_k = \la_k(\si) P_k
\ee
 (in the spirit of the adiabatic
hypothesis).  Let the operators $\wh A_m(z)$ be time-independent and preserve
the eigenspaces
$E_k$ of the Hamiltonian $\cH$, i.e.,
\beq
\wh A_m(z)= \op\sum_k \wh A_m^k(z)\circ P_k, \label{+361}
\eeq
where $\wh A_m^k(z)$, $z\in Z$, are self-adjoint operators in $E_k$.
It follows that $\wh A_m(\si)$ commute with
$\cH(\si)$ at all points of the parameter bundle $\Si\to\bR$. Then, 
restricted to
each subspace $E_k$, the parallel transport operator (\ref{+359}) is a
unitary operator in $E_k$. In this case, the Berry connection 
(\ref{+357})  on the
$U(E)$-principal bundle
$P\to Z$ can be seen as a composite connection on the composite bundle 
\be
P\to P/U(n)\to Z,
\ee
which is defined by some principal connection on the $U(n)$-principal bundle
$P\to P/U(n)$ and the trivial connection on the fiber bundle $P/U(n)\to Z$.
The typical fiber of $P/U(n)\to Z$ is exactly the classifying space $B(U(n))$
for $U(n)$-principal bundles.
Moreover, one can consider the parallel transport
along a curve in the bundle $P/U(n)$. In this case, a state vector
$\psi(t)$ acquires a geometric phase factor in addition to the dynamical
phase factor. In particular, if $\Si=\bR$ (i.e., classical parameters are
absent and  Berry's phase has only the geometric origin) we come to the case
of a Berry connection on the
$U(n)$-principal bundle on the classifying space $B(U(n))$ \cite{bohm}.
If $n=1$, this is the variant of Berry's geometric phase of Ref. \cite{anan}.

\end{document}